\documentclass[notitlepage]{article}
\usepackage{natbib}
\usepackage{graphicx}

\begin{document}

\title{Simulating radio emission from air showers with CoREAS}


\author{T. Huege\\
  \small IKP, Karlsruhe Institute of Technology (KIT)
  \and
  M. Ludwig\\
  \small EKP, Karlsruhe Institute of Technology (KIT)
  \and C.~W. James\\
  \small ECAP, University of Erlangen-N\"urnberg
}

\maketitle

\begin{abstract}
CoREAS is a Monte Carlo code for the simulation of radio emission from extensive air showers. It implements the endpoint formalism for the calculation of electromagnetic radiation directly in CORSIKA. As such, it is parameter-free, makes no assumptions on the emission mechanism for the radio signals, and takes into account the complete complexity of the electron and positron distributions as simulated by CORSIKA. In this article, we illustrate the capabilities of CoREAS with simulations carried out in different frequency ranges from tens of MHz up to GHz frequencies, and describe in particular the emission characteristics at high frequencies due to Cherenkov effects arising from the varying refractive index of the atmosphere.
\end{abstract}


\section{The CoREAS code}

Simulations of radio emission from cosmic ray air showers based on CORSIKA \citep{HeckKnappCapdevielle1998} have a long history. REAS2 \citep{HuegeUlrichEngel2007a} calculated the radio emission with the geosynchrotron approach on the basis of CORSIKA-derived multi-dimensional histograms describing the electron- and positron-distributions in air showers. Later, it became clear that the emission model in REAS2 (and almost all other time-domain models) was incomplete \citep{HuegeLudwigScholtenARENA2010}. We developed the endpoint formalism \citep{JamesFalckeHuege2010} and implemented it in REAS3 \citep{LudwigHuege2010} to provide a complete and model-free calculation of air shower radio emission. The air shower model, however, was unchanged between REAS2 and REAS3 in that it still rested on the histogrammed particle distributions. With REAS3.1, we included the effects of a realistic refractive index in air \citep{LudwigIcrc2011}.

While REAS3.1 already provided a very detailed calculation of air shower radio emission, the histogramming step between the air shower and radio simulations had some drawbacks. Correlations between parameters stored in the same four-dimensional histograms were retained, but correlations between parameters stored in different histograms were lost. This affected in particular correlations between the lateral distance of particles from the shower axis and their momentum angle with respect to the shower axis. Other characteristics that were lost during histogramming were the predominant outward drift of particles from the shower axis and the systematic offset of the electron and positron distributions (i.e., a dipolar component). Also, histogramming always smoothed the distributions and thus suppressed local over- and underdensities arising from sub-showers. Finally, the histogramming step consumed computation time and made the handling somewhat cumbersome. While histogramming was a very valuable tool during the diagnosis and development stage, it is clear that now that the fundamentals are well-understood, this intermediate step should be eliminated.

For this reason, we have built the endpoint formalism directly into CORSIKA. This new, integrated code is named CoREAS and currently is available in version 1.0. The integration is very natural because CORSIKA provides the position, time and energy of each tracked particle at the start- and endpoint of individual track segments. These are then used directly to calculate the radio emission with the endpoint formalism. Refractive index effects are taken into account correctly. Special care has to be taken near the Cherenkov angle where the endpoint formalism can diverge \citep{JamesARENA2012,BelovARENA2012}. In these situations, we have to explicitly take into account the finite time resolution of our detector by smearing out contributions on the sampling time-scale and reverting to the same far-field approximation as employed in the ZHS algorithm \citep{ZasHalzenStanev1992}. The calculation speed of CoREAS is much faster than that of REAS3, up to a factor of 10 depending on the configuration of observer positions.

CoREAS 1.0 is a mature code that has been tested thoroughly and is already in use in many experiments (LOPES, LOFAR, AERA, RASTA, ANITA, CROME). Comparisons with LOPES data have been presented at this conference \citep{LudwigARENA2012}. We plan to openly release the CoREAS source code as part of the pending release of CORSIKA version 7. In the following, we present some results derived with CoREAS at low frequencies (tens of MHz as typical for LOPES, LOFAR, AERA), high frequencies (hundreds of MHz as typical for ANITA) and very high frequencies (GHz frequencies as typical for CROME). As the performance of the code is much better than that of REAS3, we can do calculations on very fine grids and thus produce some particularly instructive plots.

\section{Simulations at low frequencies}

\begin{figure}[h!t]
\begin{minipage}{\textwidth}
  \includegraphics[height=0.19\textheight]{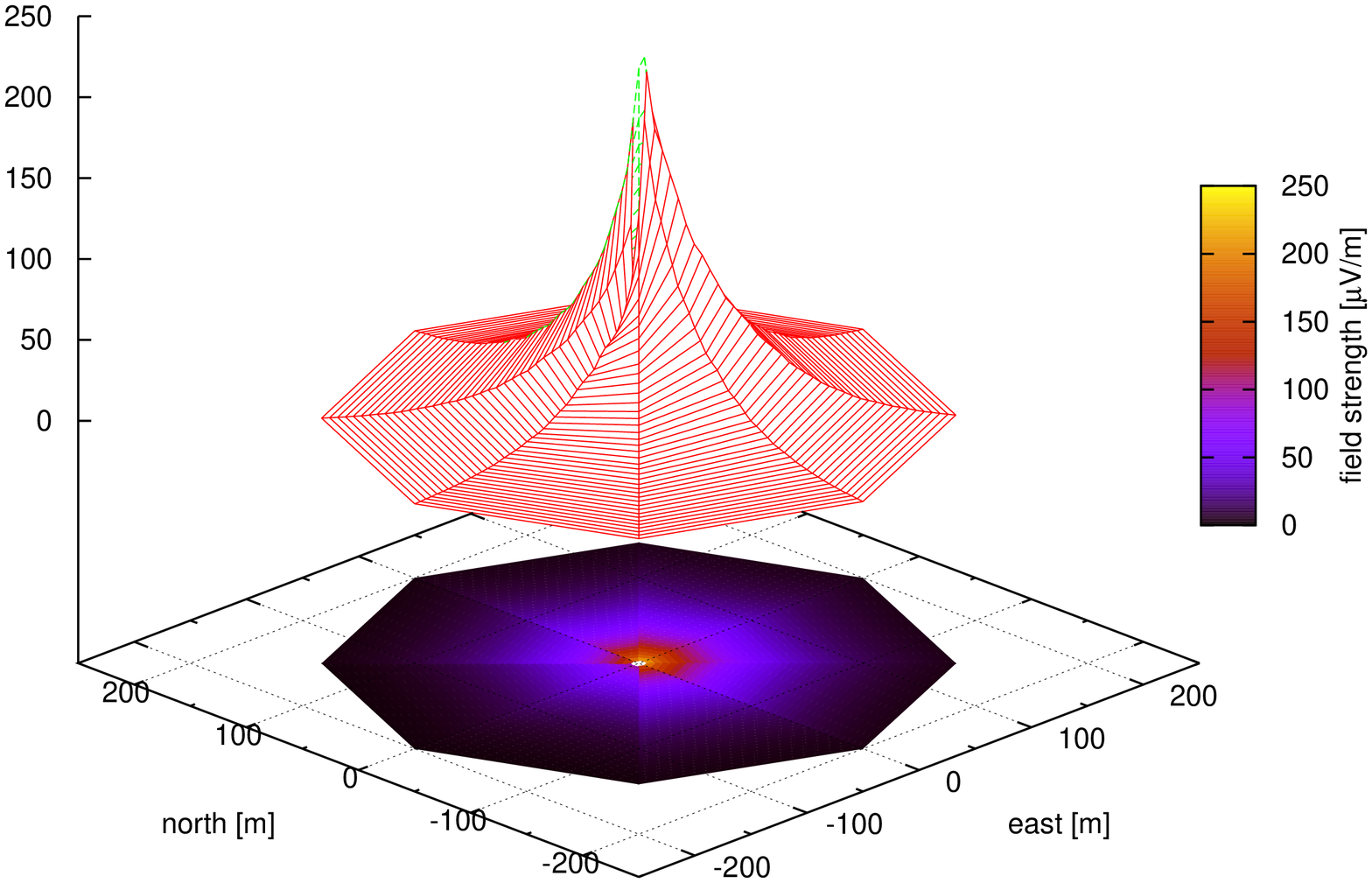}
  \hspace{0.5cm}
  \includegraphics[height=0.19\textheight]{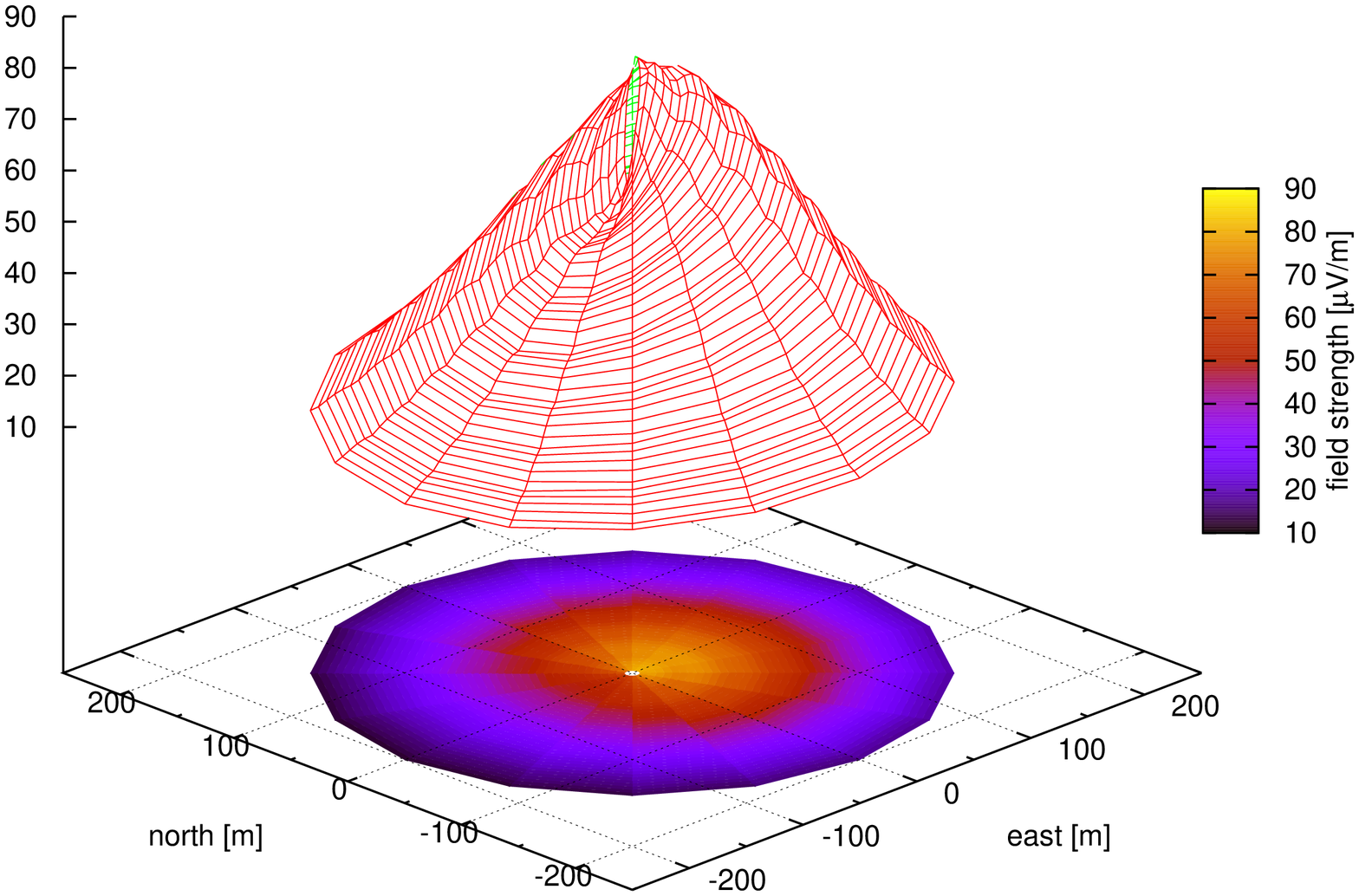}
\end{minipage}
  \caption{Footprints of the 40-80~MHz total field strength for vertical $10^{17}$~eV air showers induced at the LOPES site by a proton (left) and an iron (right) primary. Please note the different absolute scaling. Both the asymmetry of the footprint and the systematically different slope of the lateral distribution caused by the different depths of shower maximum are obvious.} \label{fig:lopesprotoniron}
\end{figure}

The main motivation for the development of modern radio emission theory were the pioneering experiments LOPES and CODALEMA working at frequency ranges of tens of MHz. A lot of progress has been made in recent years in the study of the air shower radio emission, and simulations have always been a fundamental guideline for interpreting the measurements. One of the most important goals for radio detection is to reliably reconstruct the energy and mass of the primary cosmic rays. Figure \ref{fig:lopesprotoniron} illustrates two important aspects that such analyses have to take into account: First, the footprint of the radio signal total field strength exhibits significant asymmetries. They result from the well-understood superposition of the dominant geomagnetic and sub-leading charge excess components of the radiation. In particular when fitting a lateral distribution function to radio measurements, these asymmetries have to be taken into account. Second, it becomes obvious immediately that there can be very significant differences between the lateral distribution functions of radio signals emitted by proton-induced air showers (deeper shower maximum) and iron-induced air showers (shallower shower maximum). Naturally, shower-to-shower fluctuations wash out these signatures, yet they can still be exploited in practice \citep{HuegeUlrichEngel2008,PalmieriARENA2012}.

\begin{figure}[h!t]
  \includegraphics[height=0.75\textwidth]{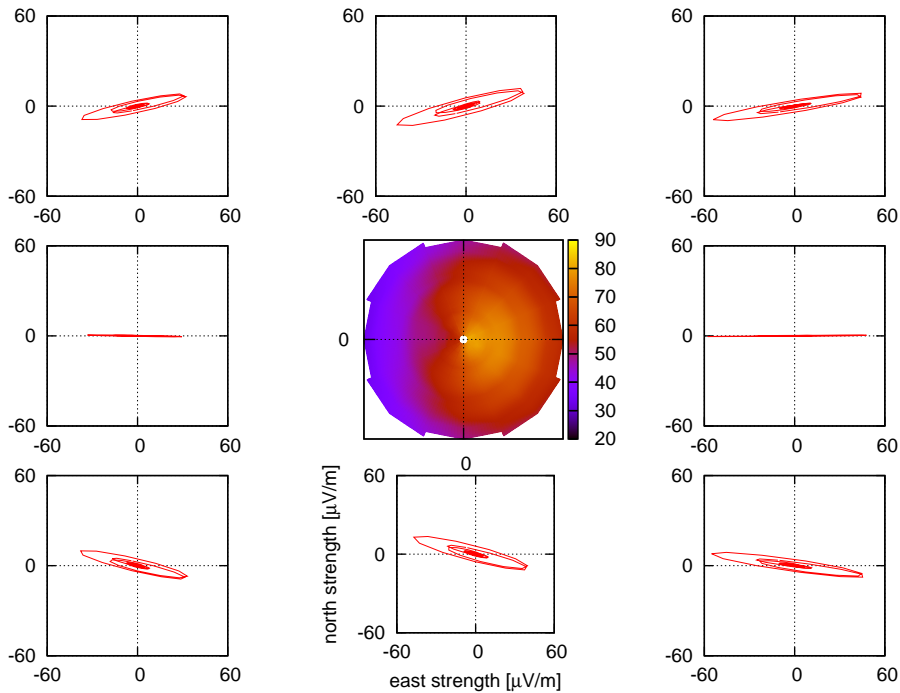}
  \caption{The time-dependence of the electric field vector in the 40-80~MHz frequency band for observers at 100~m lateral distance to the north, north-east, east , south-east, south, south-west, west, and north-west (counting clockwise from the top panel) of the axis of a vertical $10^{17}$~eV air shower induced by an iron nucleus at the LOPES site. The map in the center shows the total amplitude footprint.} \label{fig:lopespolarisation}
\end{figure}

The interaction of the geomagnetic and charge excess components of the radio emission from a vertical air shower are illustrated in some more depth in Figure \ref{fig:lopespolarisation}. In the middle, the inner 100~m radius of the radio footprint are shown. At the outside, scatter plots of the north-south component versus the east-west component of the electric field vector as a function of time as observed at various observer positions at 100~m radius illustrate the polarisation characteristics of the radio signal. For observers in the east, the geomagnetic and charge excess components superpose constructively, the resulting polarisation is purely linear east-west. Similarly, for observers in the west, the two components interfere destructively. This is the reason for the asymmetry already observed in Figure \ref{fig:lopesprotoniron}. At the other observer positions, the situation is more complex. The geomagnetic component is still east-west-polarized, yet the charge excess component has a linear polarization with the electric field vector aligned radially. In addition, the pulses caused by the two components are not completely in synch. Therefore, the superposition of the two components leads to an ellipse in the scatter plots (rather than a line). In other words, the total polarisation has a certain amount of circular polarization, and the resulting total polarisation is elliptical. These subtleties have to be kept in mind when analyzing measurements with respect to polarisation characteristics.

\section{Simulations at high frequencies}

With the inclusion of a realistic refractive index in the air shower radio emission simulations, Cherenkov effects arise for suitable geometries. This means that signals emitted at different times arrive simultaneously at an observer, or putting it differently, the emission is compressed in time. The time-compression of the signal means that the frequency spectrum expands to much higher frequencies than for a calculation with a refractive index of unity. At frequencies well above 100~MHz this can have dramatic effects.

\begin{figure}[h!t]
  \includegraphics[height=0.25\textheight]{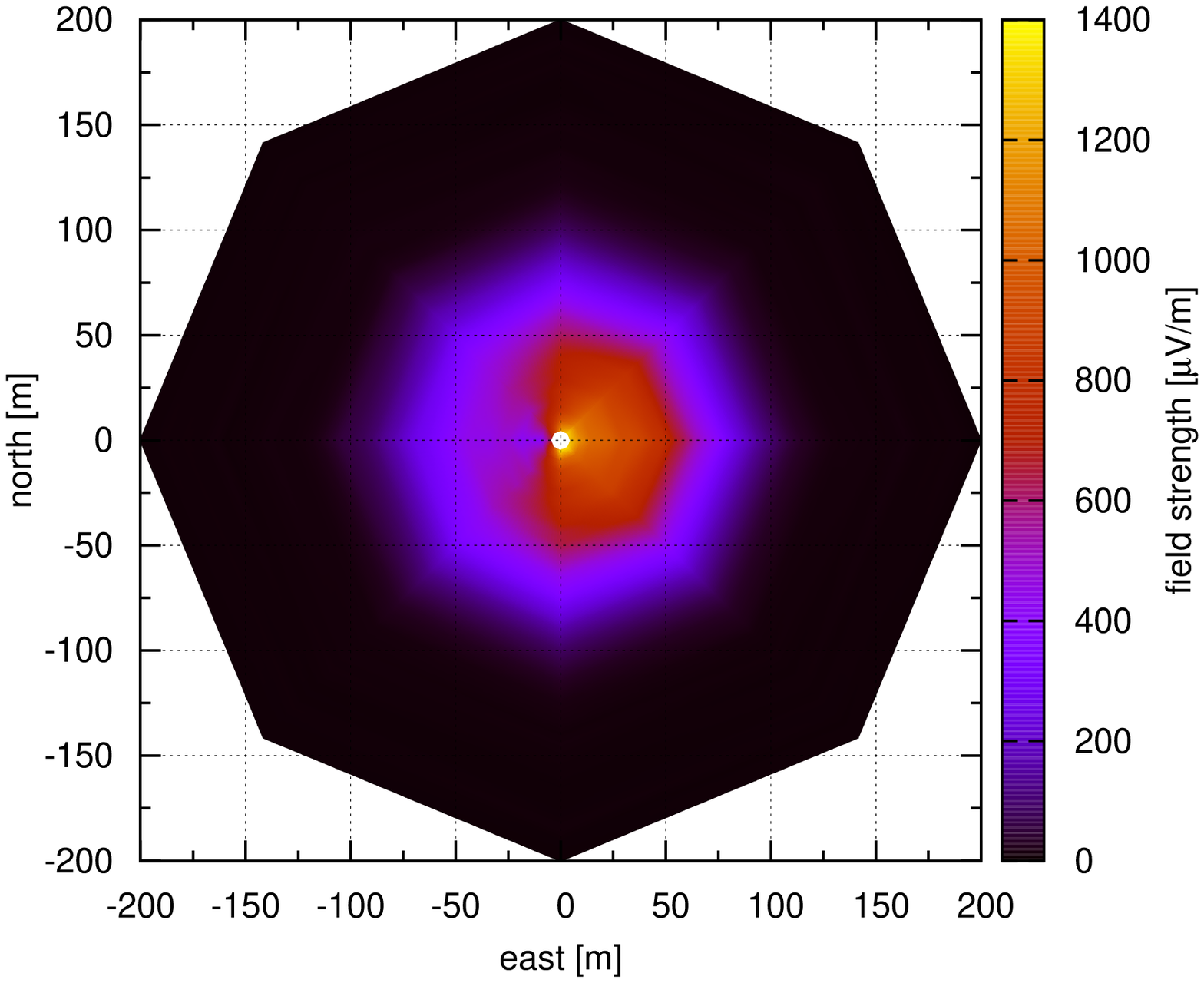} \hspace{0.5 cm}
  \includegraphics[height=0.25\textheight]{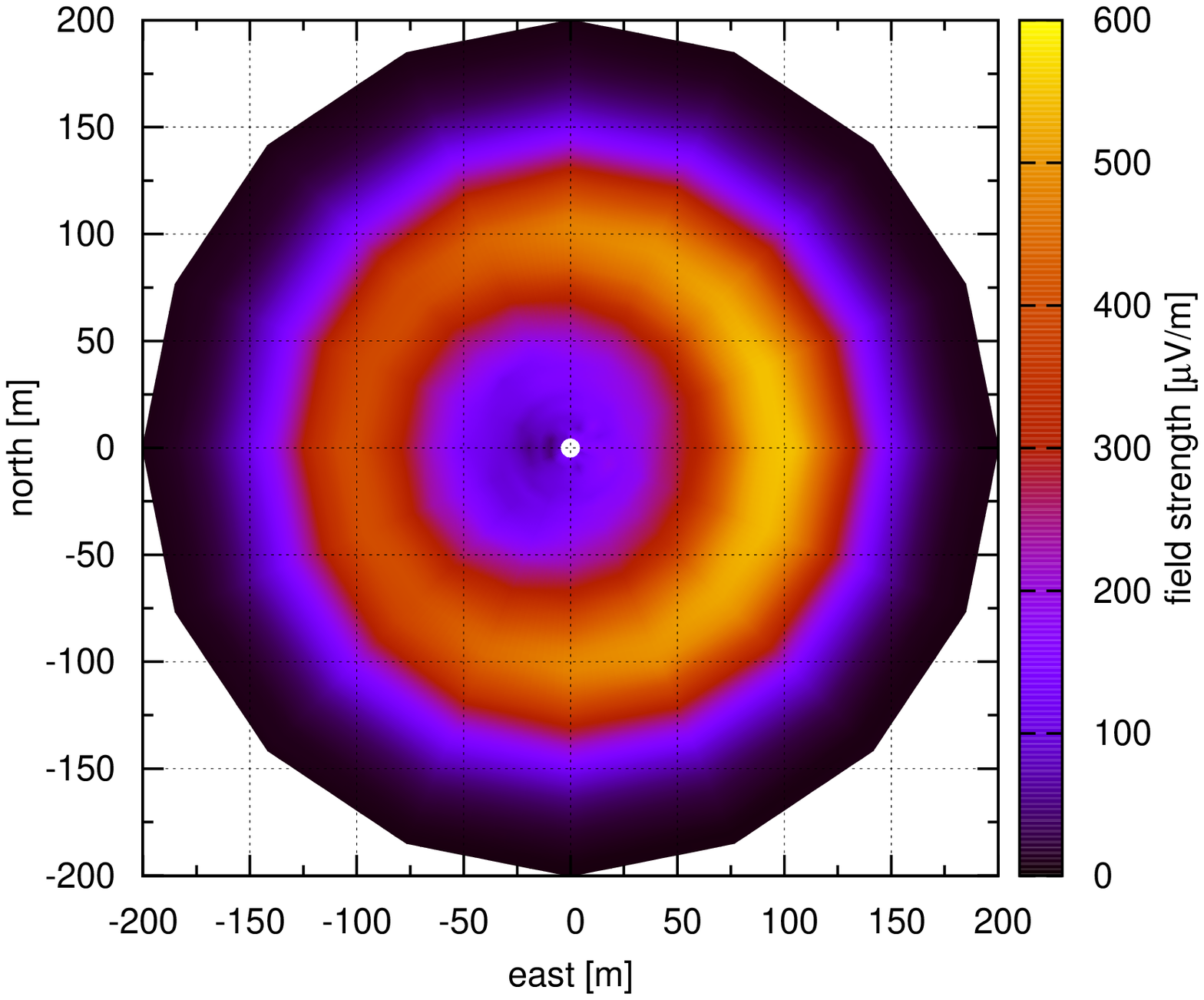}
  \caption{Footprints of the total field strength of a vertical $10^{17}$~eV air shower induced by a proton (left) and an iron (right) primary at the LOPES site as seen at 300-1200~MHz.} \label{fig:anitaprotoniron}
\end{figure}

An example illustrating these effects is given in Figure \ref{fig:anitaprotoniron}. While for the proton shower (left panel) the footprint looks similar as for low frequencies, for the iron shower (right panel) a prominent ring apppears. The ring appears for observers for which Cherenkov-like compression of the radio pulses occurs, leading to a strong signal even in the frequency range of 300-1200~MHz. The difference between the proton and iron case is the depth of shower maximum, and consequently the distance of the ``radio source'' from the observer. The proton shower penetrated deeply, and the shower maximum was close to the ground, whereas the iron shower reached its maximum earlier in the atmosphere, allowing for the Cherenkov condition to be fulfilled for observers at the adequate lateral distance. It is obvious that the diameter of the ring thus provides information on the depth of shower maximum, although it should be kept in mind that it probes the geometrical distance from source to observer, not the atmospheric depth traversed. Closer inspection of the ring also illustrates that the well-known east-west asymmetry is still present. In other words, also at these high frequencies the superposition of geomagnetic and charge excess emission is present. Overall, the polarization of the signal is analogous to the one observed at lower frequencies. For inclined air showers, the ring will become elliptical, and as for inclined air showers the source moves geometrically farther away, the ring diameter will increase.

The frequency range illustrated here is the one probed by the ANITA experiment. Comparisons have shown that the cosmic-ray events observed by ANITA can be plausibly explained by Cherenkov-compressed air shower radio emission, and studies to determine the energy scale from CoREAS simulations are ongoing \citep{BelovANITAARENA2012}.

\section{Simulations at very high frequencies}

\begin{figure}[h!t]
  \includegraphics[height=0.25\textheight]{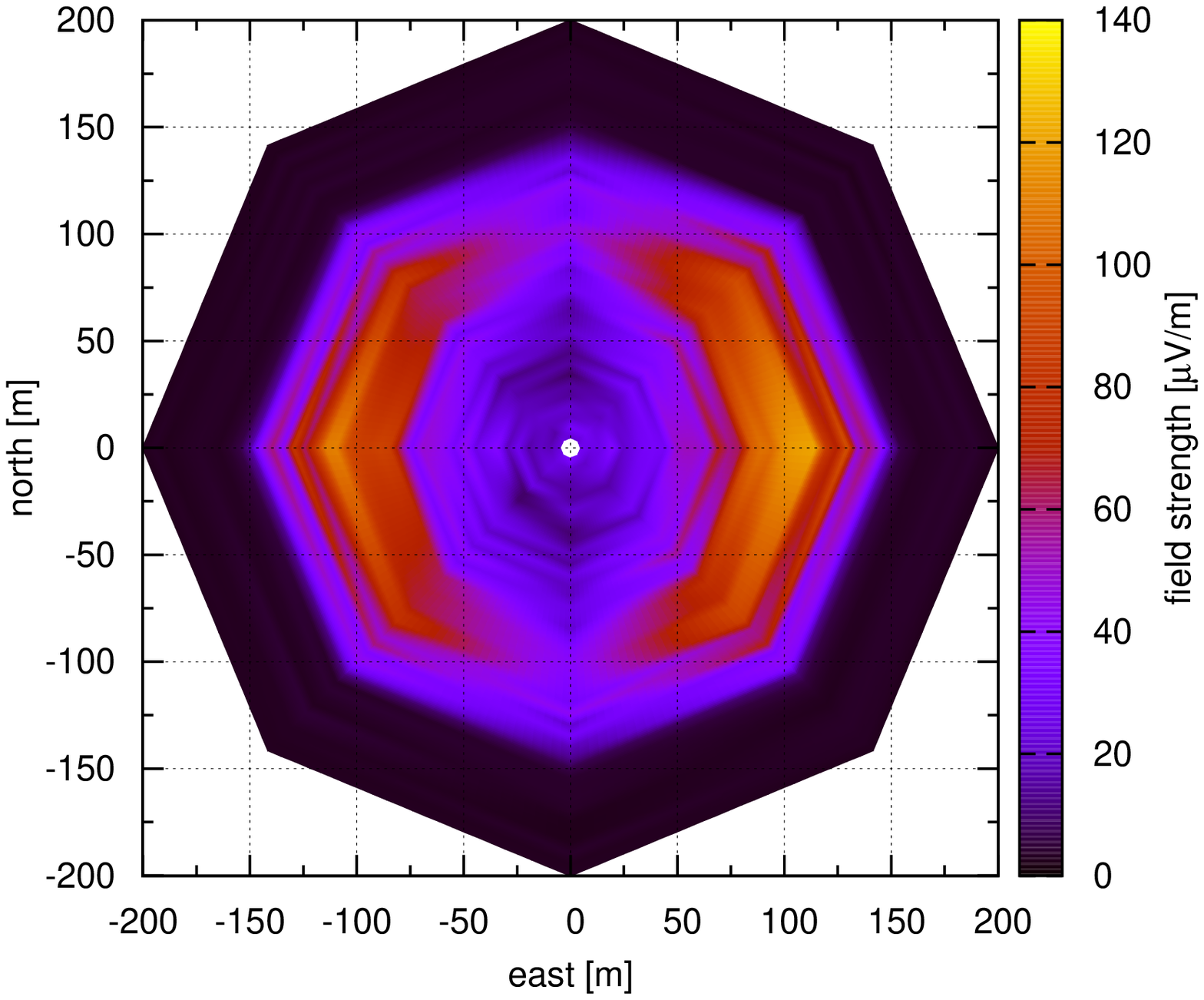} \hspace{0.5 cm}
  \includegraphics[height=0.25\textheight]{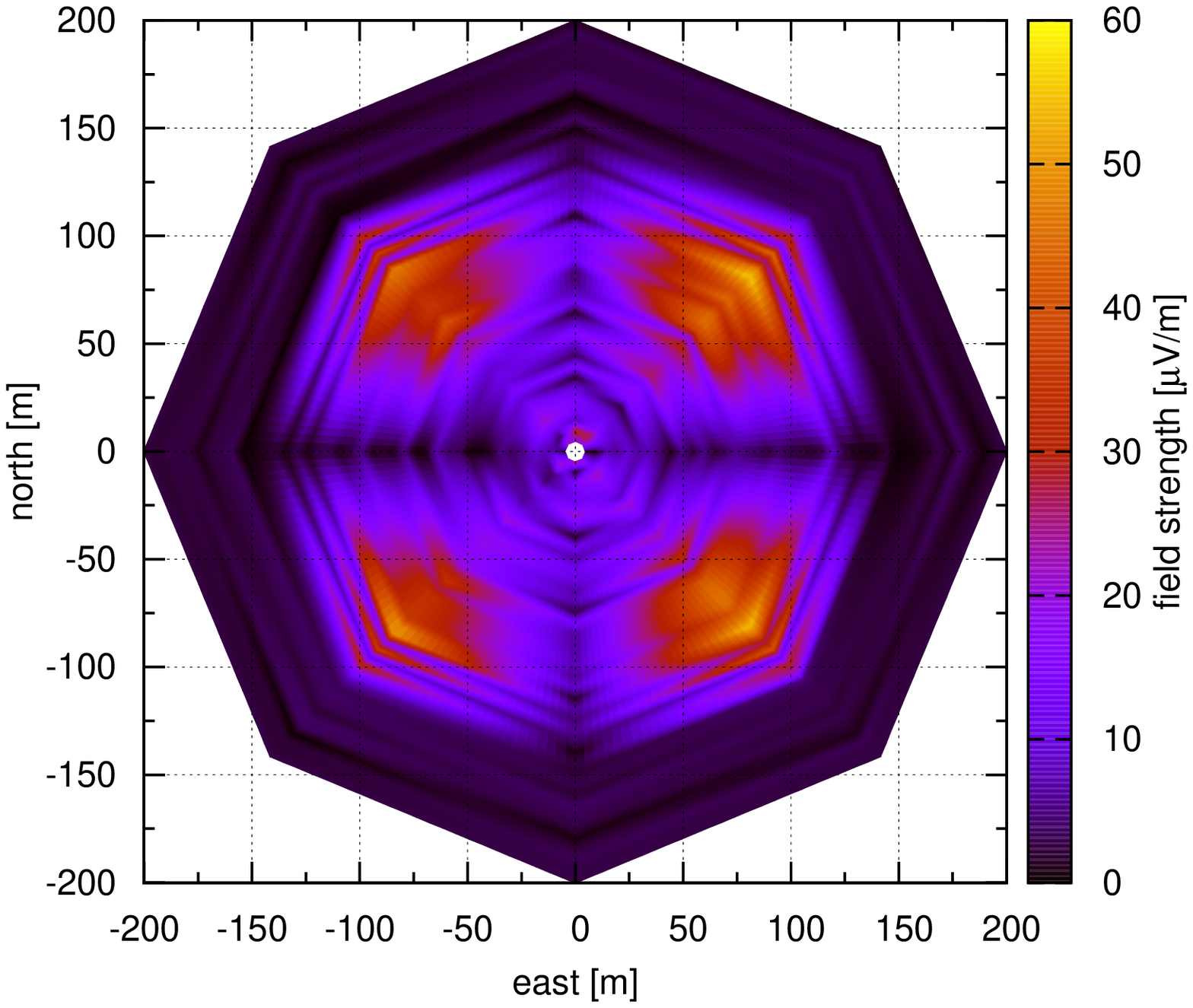}
  \caption{Footprints of the total field strength (left) and north-south component of the electric field vector (right) for a vertical $10^{17}$~eV air shower induced by a proton at the LOPES site as seen at 3400-4200~MHz.} \label{fig:crome}
\end{figure}

If one goes to yet higher frequencies such as 3.4-4.2~GHz as probed by the CROME experiment, the emission characteristics change. This is illustrated in Figure \ref{fig:crome}. While the Cherenkov ring is still clearly visible, and also an east-west asymmetry is still present, the ring for the total electric field (left panel) now has a gap for the north and south observer locations. This is in strong contrast to the emission pattern at MHz and hundreds of MHz frequencies. Likewise, if one looks at the north-south component of the electric field vector (right panel), the emission pattern shows four distinct quadrants. At lower frequencies, the footprint shows emission in the northern and southern region of the footprint, but there is no gap along the north-south axis. Another difference is that for observers in these four quadrants, the electric field vector is aligned radially with respect to the shower axis, unlike at lower frequencies (cf.\ Figure \ref{fig:lopespolarisation}). The ``clover-leaf'' emission pattern is related to the geomagnetic emission - if the magnetic field is switched off, the amplitude drops to a level of only 15-20~$\mu$V/m, and the resulting footprint of the charge excess emission is circularly symmetric. Another interesting feature is the presence of ``ripples'' in the footprint. These are not numerical artifacts, but are indeed interference minima and maxima appearing in the simulated signal.

The ``clover-leaf'' pattern visible for the north-south component of the emission at GHz frequencies is reminiscent of the patterns observed for calculations based on the geosynchrotron approach \citep{HuegeUlrichEngel2007a}. Geosynchrotron emission in this context relates to the radiation associated with the direct acceleration of the particles in the geomagnetic field rather than the emission associated with the effective drift of particles in the time-varying transverse currents arising from an equilibrium of acceleration by the magnetic field and deceleration due to interactions. A hypothesis that is still to be tested is that the geosynchrotron component actually could be relevant at such high frequencies, while it is insignificant at lower frequencies.

A first comparison of CoREAS simulations with CROME measurements \citep{WernerARENA2012} shows a general agreement, so that also at these very high frequencies radio emission from cosmic ray air showers seems to be observable.

\section{Conclusions}

With CoREAS 1.0, a full-fledged, parameter- and model-free Monte Carlo simulation code for radio emission from air showers on the basis of the endpoint formalism implemented in CORSIKA is available. It is more precise, much faster and much easier to handle than its predecessor REAS3.1, and since it uses identical input and output formats, transition to CoREAS will be easy for users who have previously used REAS for simulations. Open publication of the source code is planned for the release of CORSIKA version 7.

While the simulation code has initially been developed for application at frequencies of tens of MHz, it can also be applied at much higher frequencies. The Cherenkov effects arising from the varying refractive index of the atmosphere lead to Cherenkov compression of the emitted radio pulses and thus the appearance of Cherenkov rings in the footprints, at which emission can be observed at frequencies of several hundreds of MHz and even at GHz frequencies. The diameter of these Cherenkov rings is related to the geometrical distance of radio source and observer, and thus contains information on the depth of shower maximum.

The emission at frequencies of hundreds of MHz shows the same polarization characteristics and symmetries as it does at tens of MHz. Comparisons with measurements of the ANITA experiment are ongoing, and Cherenkov-compressed air shower radio emission is a very plausible candidate to explain the observed events.

At GHz frequencies, the radio emission from air showers shows qualitatively different polarization characteristics than at lower frequencies. Along the north-south axis from the core, the emission is attenuated, both in the total signal and in the north-south electric field component. The latter exhibits a distinct ``clover-leaf'' pattern which is associated with the magnetic field, and which was previously observed in geosynchrotron simulations. Further studies should clarify the source and nature of this pattern.





\bibliographystyle{aipproc}   


\IfFileExists{\jobname.bbl}{}
 {\typeout{}
  \typeout{******************************************}
  \typeout{** Please run "bibtex \jobname" to optain}
  \typeout{** the bibliography and then re-run LaTeX}
  \typeout{** twice to fix the references!}
  \typeout{******************************************}
  \typeout{}
 }

\end{document}